\documentclass[numberedappendix,apj,twocolumn]{emulateapj}
\usepackage{graphicx}
\usepackage{amssymb,amsmath}

\begin{document}

\title{Luminous and High Stellar Mass Candidate Galaxies at $z\approx 8$
Discovered in The Cosmic Assembly Near-Infrared Deep Extragalactic Legacy Survey
}

\author{Haojing Yan \altaffilmark{1}, 
Steven L. Finkelstein \altaffilmark{2,21},
Kuang-Han Huang \altaffilmark{3},
Russell E. Ryan \altaffilmark{4},
Henry C. Ferguson \altaffilmark{4},
Anton M. Koekemoer \altaffilmark{4},
Norman A. Grogin \altaffilmark{4},
Mark Dickinson \altaffilmark{5},
Jeffrey A. Newman \altaffilmark{6},
Rachel S. Somerville \altaffilmark{7},
Romeel Dav\'{e} \altaffilmark{8},
S. M. Faber \altaffilmark{9},
Casey Papovich \altaffilmark{10},
Yicheng Guo \altaffilmark{11},
Mauro Giavalisco \altaffilmark{11},
Kyoung-soo Lee \altaffilmark{12},
Naveen Reddy \altaffilmark{13},
Asantha R. Cooray \altaffilmark{14},
Brian D. Siana \altaffilmark{13},
Nimish P. Hathi \altaffilmark{15},
Giovanni G. Fazio\altaffilmark{16},
Matthew Ashby \altaffilmark{16},
Benjamin J. Weiner \altaffilmark{17},
Ray A. Lucas \altaffilmark{4},
Avishai Dekel \altaffilmark{18},
Laura Pentericci \altaffilmark{19},
Christopher J. Conselice \altaffilmark{20},
Dale D. Kocevski \altaffilmark{9},
Kamson Lai \altaffilmark{9}
}

\altaffiltext{1}{Department of Physics \& Astronomy, University of Missouri,
     Columbia, MO 65211, USA}
\altaffiltext{2}{Department of Astronomy, The University of Texas at Austin, 
     Austin, TX 78712, USA}
\altaffiltext{3}{Department of Physics \& Astronomy, Johns Hopkins University, 
     3400 N. Charles Street, Baltimore, MD, USA, 21218}
\altaffiltext{4}{Space Telescope Science Institute, 3700 San Martin Drive,
     Baltimore, MD 21218, USA}
\altaffiltext{5}{National Optical Astronomy Observatory, 950 N. Cherry Ave., 
     Tucson, AZ 85719, USA}
\altaffiltext{6}{Department of Physics and Astronomy, University of Pittsburgh,
     3941 O'Hara Street, Pittsburgh, PA 15260, USA}
\altaffiltext{7}{Physics \& Astronomy Department, Rutgers University, 
     Piscataway, NJ 08854, USA}
\altaffiltext{8}{Astronomy Department, The University of Arizona, Tucson,
     AZ 85721, USA}
\altaffiltext{9}{University of California Observatories/Lick Observatory, 
     University of California, Santa Cruz, CA 95064, USA}
\altaffiltext{10}{George P. and Cynthia Woods Mitchell Institute for 
    Fundamental Physics and Astronomy, and Department of Physics and Astronomy,
    Texas A\&M University, College Station, TX 77843-4242, USA}
\altaffiltext{11}{Department of Astronomy, University of Massachusetts, Amherst,
     MA 01003, USA}
\altaffiltext{12}{Department of Physics, Purdue University, 525 Northwestern 
     Ave., West Lafayette, IN 47907, USA}
\altaffiltext{13}{Department of Physics \& Astronomy, University of California, 
     Riverside, CA 92521, USA}
\altaffiltext{14}{Deptment of Physics \& Astronomy, University of California,
     Irvine, CA 92697}
\altaffiltext{15}{Observatories of the Carnegie Institution for Science, 813 
     Santa Barbara Street, Pasadena, CA 91101, USA}
\altaffiltext{16}{Harvard-Smithsonian Center for Astrophysics, 60 Garden Street,
     MS65, Cambridge, MA02138, USA}
\altaffiltext{17}{Steward Observatory, University of Arizona, 933 N. Cherry St.,
     Tucson, AZ 85721, USA}
\altaffiltext{18}{Racah Institute of Physics, The Hebrew University of 
    Jerusalem, Jerusalem 91904, Israel}
\altaffiltext{19}{INAF Osservatorio Astronomico di Roma, Via Frascati 33,00040
     Monteporzio (RM), Italy}
\altaffiltext{20}{Centre for Astronomy and Particle Theory, University of 
     Nottingham, University Park, Nottingham, NG7 2RD, UK}
\altaffiltext{21}{Hubble Fellow}

\begin{abstract}

  One key goal of the {\it Hubble Space Telescope} Cosmic Assembly 
Near-Infrared Deep Extragalactic Legacy Survey is to track galaxy
evolution back to $z\approx 8$. Its two-tiered ``wide and deep'' 
strategy bridges significant gaps in existing near-infrared surveys. Here
we report on $z\approx 8$ galaxy candidates selected as F105W-band
dropouts in one of its deep fields, which covers 50.1~arcmin$^2$ to 4~ks depth
in each of three near-infrared bands in the Great Observatories Origins Deep
Survey southern field. Two of our candidates have $J<26.2$~mag, and are 
$> 1$~mag brighter than any previously known F105W-dropouts. We derive 
constraints on the bright-end of the rest-frame ultraviolet luminosity function
of galaxies at $z\approx 8$, and show that the number density of such very
bright objects is higher than expected from the previous Schechter luminosity
function estimates at this redshift. Another two candidates are securely
detected in {\it Spitzer} Infrared Array Camera images, which are the first
such individual detections at $z\approx 8$. Their derived stellar masses are on
the order of a few $\times 10^9$~$M_\odot$, from which we obtain the first
measurement of the high-mass end of the galaxy stellar mass function at 
$z\approx 8$. The high number density of very luminous and very massive
galaxies at $z\approx 8$, if real, could imply a large stellar-to-halo
mass ratio and an efficient conversion of baryons to stars at such an early
time. 

\end{abstract}

\keywords{cosmology: observations --- galaxies: luminosity function, mass function}

\section{Introduction}

  In the recent years, deep near-infrared imaging surveys have begun to yield a
significant number of candidate galaxies at very high redshifts. The wide-field
surveys from the ground have produced a handful of bright candidates at 
$z\approx 7$ (e.g., Ouchi et al. 2009; Hickey et al. 2010; Castellano et al.
2010; Capak et al. 2011; Hsieh et al. 2012; Hathi et al. 2012), while
the pencil-beam survey by the {\it Hubble Space Telescope} (HST) Wide Field 
Camera 3 (WFC3) within the historical Advanced Camera for Surveys (ACS) Hubble
Ultra Deep Field (hereafter ACS-HUDF; Beckwith et al. 2006) and its two parallel
fields (HUDF09; PI. Illingworth) have probed the fainter populations at
$z\approx 7$--8 (Oesch et al. 2010; Bouwens et al. 2010, B10; Bunker et al.
2010; McLure et al. 2010, M10; Yan et al. 2010, Y10; Finkelstein et al. 2010)
and possibly out to $z\approx 10$ (Yan et al. 2010; Wyithe et al. 2011;
Bouwens et al. 2011a).
The WFC3 Early Release Science program (ERS, PI. O'Connell; Windhorst et al.
2011) has played an important role in connecting the ``wide'' and the ``deep''
ends of the exploration, allowing construction of $z\approx 7$--8 samples at
intermediate brightness levels (Wilkins et al. 2011; B11; Lorenzoni et al. 
2011, L11). Meanwhile, the Hubble Infrared Pure Parallel Imaging Extragalactic
Survey (HIPPIES; Yan et al. 2011) and the Brightest of Reionizing Galaxies
Survey (Trenti et al. 2011a) have been exploring the bright end of the
population through a large number of random, discrete WFC3 pointings obtained
during the {\it HST} parallel orbits. Surveys in foreground cluster fields
utilizing gravitational lensing magnification have also resulted in a handful
of promising candidates at $z\gtrsim 7$--9 (e.g., Richard et al. 2006; 
Bradley et al. 2008, 2012; Laporte et al. 2011; Hall et al. 2012), which
complement the surveys in blank-sky fields.

  The Cosmic Assembly Near-Infrared Deep Extragalactic Legacy Survey
(CANDELS; PIs: Faber \& Ferguson; see Grogin et al. 2011 and Koekemoer et al.
2011) employs a two-tiered, ``wide and deep'' strategy, which makes it
uniquely positioned in bridging the significant gaps among existing surveys.
In particular, its ``Deep'' component will cover $\sim 125$~arcmin$^2$ in two
fields upon completion, and its data will be ideally suited for studying
galaxies at very high redshifts. In this paper, we report our preliminary
results from the Deep observations in the Great Observatories Origins Deep
Survey (GOODS; Giavalisco et al. 2004) southern field, where we use the nearly
complete data set to study the galaxy population at $z\approx 8$. 
We use the following cosmological parameters throughout:
$\Omega_M=0.27$, $\Omega_\Lambda=0.73$ and $H_0=71$~km~s$^{-1}$~Mpc$^{-1}$. The
quoted magnitudes are all in the AB system.

\section{Data and Photometry}

   As detailed in Grogin et al. (2011), the CANDELS/Deep region in the GOODS-S
field is located in its middle stripe where we have the deepest GOODS
{\it Spitzer} coverage, which also includes the ACS-HUDF. The data used in this
study include 100\% of the observations in F105W (hereafter $Y_{105}$) and 
83\% of the observations in F125W ($J_{125}$) and F160W ($H_{160}$) scheduled
for this field. The typical exposure times per sky position are 8090, 7450, and
7770 seconds in $Y_{105}$, $J_{125}$ and $H_{160}$, respectively. Our present
analysis is confined to the region where the effective integration times are at
least 4000 seconds in all three WFC3 bands, which is 62.9~arcmin$^2$ in size.
For the reason described below, we further exclude the region covered
by the ACS-HUDF data, and the final effective area in our current study is 
50.1~arcmin$^2$.

    The products of the CANDELS data reduction (Koekemoer et al. 2011) include 
the science mosaics and the associated ``root mean square'' (RMS) maps. These 
RMS maps account for correlations between pixels that are 
introduced during the data reduction procedures (see e.g., Dickinson et al.
2004). Such correlations tend to reduce the measured pixel-to-pixel background
noise in the {\it HST} images, and can lead photometry programs to
underestimate the actual photometric uncertainties.  We avoid this problem by
computing uncertainties using the normalized RMS maps, summing their values in
quadrature over the measurement apertures, and including an additional term for
Poisson noise from the source flux (generally negligible for very faint dropout
galaxies). The values of the CANDELS RMS maps have been validated both by
comparison with autocorrelation measurements of the image noise, and by
analysis of the dispersion of measurements in apertures randomly distributed
within empty regions of the images. Our final mosaics have a pixel scale of
$0\farcs 06$, and are tied to the same astrometric frame as the ACS data in
GOODS. 

  The GOODS ACS data are used in our study to confirm the optical
non-detections of the candidates (see \S 3.1). These data have been reprocessed
to make mosaics at the same $0\farcs 06$ pixel scale for CANDELS. On average,
the 5-$\sigma$ sensitivities (as measured from the RMS maps) within a 
$0\farcs 2$-radius aperture in ACS F435W, F606W, F775W, F850LP (hereafter 
$B_{435}$, $V_{606}$, $i_{775}$, $z_{850}$) and WFC3 $Y_{105}$, $J_{125}$, and
$H_{160}$ are 28.04, 28.25, 27.65, 27.48, 28.15, 28.05, and 27.82~mag,
respectively. 

  The HUDF region presents a complexity in our analysis. Both the HUDF09 WFC3
IR data and the ACS-HUDF data are much deeper than their counterparts that
cover the rest (and much wider) area of interest. If we were to use these
deeper data in our candidate selection, it would be very difficult to uniformly
carry out the simulations that are necessary for the follow-up analysis
(see \S 3.4). For simplicity, we therefore exclude the ACS-HUDF region from
this current work.

   In addition to the GOODS ACS data, we also have new ACS data in F814W
(hereafter $I_{814}$) that CANDELS took in coordinated parallel mode to the
WFC3 IR observations (see Grogin et al. 2011). These new ACS data suffer 
severely from the degraded charge transfer efficiency (CTE) of the aged CCDs,
however, and the empirical correction of Anderson \& Bedin (2010) were applied
first before they were processed by the CANDELS reduction pipeline.

   The photometry in WFC3 and ACS was done together by running the SExtractor
software (Bertin \& Arnouts 1996) in dual-image mode. We used the sum of the
$J_{125}$ and the $H_{160}$ mosaics, each weighted by the inverse square
of their RMS maps, as the detection image. A 3-pixel FWHM, 7$\times$7 Gaussian
filter was used to convolve the detection image, the threshold was set to 
0.8-$\sigma$, and at least 4 connected pixels above this threshold are
required. SExtractor provides a number of methods to measure flux, among which
{\tt MAG\_APER}, {\tt MAG\_ISO} and {\tt MAG\_AUTO} are commonly used in dropout
selections. {\tt MAG\_APER} magnitudes are calculated within a circular aperture
of fixed size. As most high-redshift 
galaxies are compact, using these magnitudes can still be appropriate when the 
circular aperture size is carefully chosen so that it maximizes the S/N and 
minimizes the loss of light for the majority of the sources of interest. While
it has the advantage of being robust in extracting very faint sources, 
{\tt MAG\_APER} has the disadvantage that real galaxies, which have different
light profiles, have different fractions of light lost for a fixed aperture. 
{\tt MAG\_ISO} magnitudes are calculated within isophotal areas, which are set
by the detection threshold and hence trace the different shapes of real
galaxies. For this reason, {\tt MAG\_ISO} usually can be tuned to maximize S/N
for galaxies of different morphologies. However, it does not capture the total
light of a galaxy. {\tt MAG\_AUTO} magnitudes are calculated
using adaptive elliptical apertures based on the first moment of the light
distributions of galaxies (Kron 1980), and are the most suitable for deriving
total magnitudes if an appropriate ``Kron factor'' is chosen. However, using
a large Kron factor to capture total light usually means that the apertures
are set too large for very faint sources such that their S/N are not optimized.
Bearing both the advantages and the caveats in mind, we adopted the
{\tt MAG\_AUTO} magnitudes (using the default 
[Kron factor, minimum radius]$=$[2.5, 3.5]) in this work.

   Dropout selection uses hard boundaries in color space (see \S 3.1). As a
result, photometric errors cause objects be scattered in and out of the
selection area: an object whose intrinsic colors meet the selection threshold
could be missed because of the errors in its measured colors, and vice versa.
While this effect can be taken into account statistically by applying the 
correction for the ``effective volume'' (see \S 3.4), the candidate samples
constructed using different photometry could differ significantly on an
object-by-object basis. To demonstrate this effect, we also constructed a
separate sample using {\tt MAG\_ISO} magnitudes, which is presented in the
appendix and is compared to the {\tt MAG\_AUTO} sample.
In both cases, we only kept the sources that have S/N~$\geq$~5.0 
in at least one band among $J_{125}$ and 
$H_{160}$, measured within the {\tt MAG\_AUTO} or the {\tt MAG\_ISO} apertures.

    We also make extensive use of the GOODS {\it Spitzer} IRAC data (Dickinson
et al. 2004), which also have the same sky projection as other data here,
and have a pixel scale of $0\farcs 6$. We mostly concentrate on the $3.6$ and
$4.5$~$\mu$m channels (hereafter [3.6] and [4.5]) in this study. We used a
version of the GOODS photometry where the detection was done with a ``mexhat''
filter to optimize de-blending. We adopted {\tt MAG\_APER} within a 
$1\farcs 5$-radius (i.e., 2.5 pixels) aperture, and obtained the ``total
magnitudes'' by applying the corrections as determined in the GOODS program
where the aperture of this size has been used (e.g., Yan et al. 2005), which
are $-0.55$ and $-0.60$~mag for [3.6] and [4.5], respectively. 

\section{$Y_{105}$-dropout Selection of Candidate Galaxies at $z\approx 8$}

  In this section, we describe our samples of candidate galaxies at 
$z\approx 8$ selected as $Y_{105}$-dropouts and the possible biases in our
selections.
  
\subsection{Color Criteria}

   Our first color criterion for $Y_{105}$-dropout selection is
$Y_{105}-J_{125} > 0.80$~mag. This is the same as that employed by B10 and Y10,
and is significantly larger than that of $Y_{105}-J_{125}>0.45$~mag used in B11
and Oesch et al. (2012) where those authors have opted to included more galaxies
at lower redshifts (starting at $z\gtrsim 7.2$) in their $Y_{105}$-dropout
selection. Our larger color decrement threshold is sensitive to Lyman-break at 
$z\gtrsim 7.7$. When the objects have S/N~$<1$ in $Y_{105}$, we
use their 2~$\sigma$ upper limits and only keep those that have 
$Y_{105}-J_{125}> 0.80$~mag as calculated using these conservative upper
limits. Our second criterion is that a $Y_{105}$-dropout should have S/N~$<2$
in all four bands of GOODS ACS images ($B_{435}V_{606}i_{775}z_{850}$). This is
to implement the invisibility requirement in these ``veto'' images, which is a
necessary condition (though not sufficient) that the detected $Y_{105}-J_{125}$
color decrements in our candidates are not features in galaxies at low
redshifts (for example, the 4000\AA\, break at $z\approx$ 2) that could mimic
the Lyman-break, because such low-redshift galaxies could have detectable fluxes
in these bands. Setting the threshold to a quantitative value of S/N~$<2$ is
somewhat arbitrary, however. Based on experience, adopting S/N~$<2$ for
invisibility usually is a good compromise between minimizing the incompleteness
and keeping the size of the initial candidate sample manageable for the later
visual inspection (see below).
To exclude as much as possible the contamination from the red galaxy 
population at low redshifts, we impose a third criterion of
$J_{125}-H_{160}\leq 0.3$~mag. This also largely limits our
selection window to $z\lesssim 8.7$, and could be biased against genuine
galaxies at $z\approx 8$ that have unusually large dust reddening or
extraordinarily old stellar populations should they exist.
Figure 1 shows our selection in the $Y_{105}J_{125}H_{160}$ space.

   Each of these criteria introduces incompleteness, for which we will correct
statistically (see \S 3.4).

\subsection{Visual Inspection}

   The initial candidates have been visually inspected to ensure that they
are legitimate sources in the IR images and that there is no reason to suspect
that their photometry might be unreliable. Examples of problems that are not
accounted for in the statistical noise model (and thus require visual 
inspection) include diffraction spikes, poorly rejected cosmic rays, 
inappropriate deblending, and bad background subtraction due to nearby
neighbors. While such issues are rare in the entire photometric catalog, they
can be selected when looking for objects with unusual colors, such as dropouts.
Visual inspection step is necessary to reduce the rate of contamination, and is 
commonly adopted in dropout selections by various groups. It is somewhat 
subjective; different inspectors might not agree on a particular candidate, and
even the same inspector might not be able to completely reproduce at different
time his/her results on a large set of samples. Nevertheless, visual inspection
is irreplaceable to get a reliable sample.

   Ideally, the simulation discussed in \S 3.4 should also include this step
when deriving the incompleteness correction. However, it is impractical to
examine the 90,000 artificial galaxies inserted during the simulation. 
Nevertheless, we do not think that many of these would have been 
inappropriately rejected by visual inspectors, because the inserted galaxies
do not resemble the artifacts or those with skewed photometry that we
typically reject.  Relative to the 90,000 inserted objects, we suspect only a
handful would have these problems. Thus we do not think that the failure to
visually inspect the artificial galaxies has a significant effect on our 
computation of the incompleteness and the effective volume.  

   In our visual inspection process, the initial
candidates were first examined by one of us (HY) to get rid of the most 
obvious contaminators ($\gtrsim 300$ objects), such as image defects close
to the field edges, diffraction spikes of bright stars, bad pixels, residuals
of IR persistence, etc. The remaining candidates ($\lesssim 200$) 
were passed to six persons for independent inspections using a three-grade
system of ``firm acceptance", ``firm rejection" and ``border-line case".
To be included in the final samples, a candidate should be firmly accepted by
at least three inspectors, and should not be firmly rejected by more than one
inspector. As described below, this approach has allowed us to study the
possible systematic biases caused by visual inspections.

   We note that the contemporaneous CANDELS $I_{814}$ images have also been 
used in our current selection process. However, the flux upper limit estimate 
is subtle in these CTE-restored images when a source is undetected in single
exposures, and a rigorous treatment to be adopted by our team is still yet to
be finalized. To avoid possible changes to our sample in the future because of
this reason, at this stage we refrain from directly using the $I_{814}$ upper
limits or S/N values in the quantitative veto process, and just use the 
$I_{814}$ images in the visual inspection step (nevertheless, we do confirm
that our current final candidates indeed formally satisfy the invisibility
requirement in $I_{814}$; see \S 3.3).

  In addition to the usual examination of the mosaics, we have taken extra
steps for assurance. In the IR, we have followed Yan et al. (2011) and
examined the final candidates in their single-epoch WFC3 science images and the
associated ``data-quality'' flags to confirm that they are not due to
instrument defects. In the optical, we have constructed the ``$\chi^2$'' sum
image of the four GOODS ACS $B_{435}V_{606}i_{775}z_{850}$ bands for our 
candidates. Based on Szalay et al. (1999), the $\chi^2$ sum of these images
optimally combines the signals from different bands and extend to a fainter
limit that achieved by individual bands. The $\chi^2$ images of our final
candidates all have $\chi^2<3.73$, which is consistent with being drawn
from a random sky distribution (Szalay et al. 1999) and thus further confirms 
their non-detections in the ACS. 

\subsection{Discussion of Sample}

   The final candidates in our sample are listed in Table 1. Their images
are displayed in Figure 2.

\subsubsection{``Negative Image'' Test}

   To investigate the possible contamination of our sample by noise spikes
due to background fluctuations, we carried out the so-called negative image
test (e.g., Dickinson et al. 2004). The image mosaics described in \S 2,
which all have zero mean background by construction (see
Koekemoer et al. 2011), were multiplied by $-1$ to make any positive pixels to
negative and vise versa. SExtractor was run in dual-image mode again on these 
negative images in the same way as in \S 2, using the weighted sum of the
the negative $J_{125}$ and $H_{160}$ images as the detection image.
After applying the same selection criteria for our $Y_{105}$-dropouts as in
\S 3.1, only three ``objects'' survived, all of which have $J_{125}>27.5$~mag,
and all only show up in the negative $J_{125}$ image.
Their morphologies have odd shapes and are typical of noise spikes, and would
be easily rejected during the visual inspection step. Therefore, we conclude
that the contamination due to noise spikes is negligible in this work. 

\subsubsection{Interlopers}

   Our stringent requirements of non-detection in the ACS (individual 
bands and $\chi^2$ sum) as well as $J_{125}-H_{160}\leq 0.3$~mag are effective
in removing contaminations from red galaxies at intermediate redshifts (see
Figure 1). In addition, the IRAC data for the isolated candidates
support that they are unlikely contaminators of this kind, because
such objects, if very weak in optical, should have very red colors at
1--4~$\mu$m such that $J_{125}-[3.6]\gtrsim 2$~mag (e.g., Yan et al. 2004).
Our candidates that have IRAC flux measurements
are all at $J_{125}-[3.6]\lesssim 1.3$~mag.

   Another possible contaminant could be cool Galactic dwarf stars. Our 
criterion of $Y_{105}-J_{125}>0.80$~mag is large enough to exclude most
such contaminations. In addition, the number density of
M/L/T-dwarfs in a high galactic latitude region such as the GOODS-S is
likely negligible (Ryan et al. 2011).
Furthermore, our detailed morphological analysis in the accompanying paper
(R. Ryan et al. 2012, ApJ submitted) shows that our brightest candidates
are more consistent with extended light profiles than
point-like.

   Using the 4~Ms {\it Chandra} catalog of Xue et al. (2011), we find no X-ray
counterparts to our candidates within $6^{''}$. There is thus no 
evidence that these objects are AGN rather than normal galaxies.
  
\subsubsection{Comparison to the Sample from Oesch et al. (2012)}

   Another paper based on the similar data set as ours was submitted by an
independent group right after ours (Oesch et al. 2012, hereafter O12), and
here we compare our sample to theirs based on the latest update available to
us (P. Oesch, private communication, 2012). 

   In the overlapping region, the WFC3 IR data are essentially the same in both
studies. However, O12's reduction of the CANDELS WFC3 data is different from 
ours, and hence the end products of the science mosaics and the associated RMS
maps are different. The photometry is also different. O12's detection image is
a $\chi^2$ image derived from the $J_{125}$ and $H_{160}$ images, whereas ours
is the weighted sum of the
$J_{125}$ and $H_{160}$. While they used {\tt MAG\_AUTO} as we also did,
they adopted a smaller aperture (1.2 Kron) to compute colors. In addition,
they stacked all the available ACS data in this region. We refrained from
combining the late ACS data that suffer from the CTE problem with the earlier
GOODS V2.0 $B_{435}V_{606}i_{775}z_{850}$ ACS data, and for the same reason we
only used the $I_{814}$ data in the visual inspection step (see \S 3.2). 
Finally, O12 adopted rather generous color criteria of 
$Y_{105}-J_{125}>0.45$~mag and $J_{125}-H_{160}<0.5$~mag, while we used
$Y_{105}-J_{125}>0.80$~mag and $J_{125}-H_{160}\leq 0.3$~mag.

   All these factors contribute to the differences between the O12 sample and
ours. It is interesting and instructive to compare the inclusion of the O12
candidates in our samples, and vice versa. Based on the photometry reported in
O12, nine of their 14 $Y_{105}$-dropouts in the same field would be expected
to enter our samples. Our sample, as shown in Table 1, include five of
them. Among the four candidates not in our sample, three of them,
CANDY-2209246371, 2432246169 and 2277945141 have $S/N<5$ in $J_{125}$ and 
$H_{160}$ and are not included
in our {\tt MAG\_AUTO} catalog, and the other one, CANDY-2272447364, has 
$Y_{105}-J_{125}=0.18$ based on our photometry. 

    We note that one of our brightest candidates, 
{\tt AUTO\_100}, is not in the O12 sample. O12 believes that this 
object has a 2.4 $\sigma$ detection in $I_{814}$, derived in the conventional
way using SExtractor. While our selection does not
use the formal S/N estimates in $I_{814}$ (see \S 3.2), we have also derived
such estimates and confirmed after-the-fact that all our final candidates
formally have S/N$<2$. This particular object has S/N$<2$ in both the 
{\tt MAG\_AUTO} and the {\tt MAG\_ISO} apertures (see also appendix). We
attribute this discrepancy to the differences in photometry as outlined
in \S 3.3.2.

    To conclude, the level of overlap and discrepancy between these two samples
is all well expected given the level of differences in their constructions. We
further demonstrate this in the appendix.

\subsection{Redshift Selection Functions}

   A given choice of Lyman-break color criteria, applied to a particular data
set, leads to a probability distribution as a function of redshift and 
magnitude, $P(m,z)$ for selecting galaxies as dropout candidates. This 
probability distribution, in turn, corresponds to an effective volume 
($V_{eff}$) for Lyman-break selection as a function of magnitude 
(Steidel et al.\ 1999).  We have carried out extensive simulations to derive
$P(m,z)$ and $V_{eff}$, in order to correct for the sample incompleteness that
is inherent in the dropout technique. The results are the redshift selection
functions at different brightness levels, which are shown in Figure 3.

   To derive these functions, we put 90,000 artificial galaxies into the WFC3
and the ACS mosaics and calculated their rate of recovery $P(m,z)$ after 
applying the same color criteria as we did when constructing our dropout
samples. These artificial galaxies have different fluxes and morphologies, with
70\% being exponential disks and 30\% having the de Vaucouleurs profile. Their
sizes follow a log-normal distribution and $\theta(f)\propto f^{0.33}$, where
$\theta$ is the peak size at flux $f$ and is normalized to $0\farcs 2$ at 
$H_{160}=26$~mag. The input spectrum is from the models of Bruzual \& Charlot
(2003; BC03), and has a constant star formation history (SFH) and an age of
50~Myr. We adopted the extinction law of Calzetti (2001), and assumed that the
distribution of E(B-V) is a Gaussian function with 
mean $\langle$E(B-V)$\rangle = 0.1$~mag and scatter $\sigma = 0.1$~mag,
restricted to E(B-V)~$> 0$.
The SEDs of the simulated
galaxies were attenuated by the line-of-slight H I absorption according to
the recipe of Meiksin (2006), which is very close to that of Madau (1995).

   By calculating the ratio of the number of the recovered galaxies and that of
the input ones, we obtained the selection function for the incompleteness
correction. Figure 3 presents it as a function of $J_{125}$ {\tt MAG\_AUTO}
magnitudes, as these magnitudes are taken as the total magnitudes, and are the
closest to the input total magnitudes of the simulated galaxies.

\subsection{IRAC Counterparts}

  We have searched for IRAC counterparts of our $Y_{105}$-dropouts within 
$r=0\farcs 6$, and the results are also summarized in Table 1. Only two objects
are securely detected in IRAC, namely, {\tt AUTO\_035} and
{\tt AUTO\_293}. As we will discuss later, their WFC3 versus IRAC 
colors are very consistent with being at high redshift. 
Objects {\tt AUTO\_204, 212, 368}
are in isolated regions, and they are invisible in the GOODS IRAC data. For all
these objects, their
2-$\sigma$ flux upper limits in $[3.6]$ and $[4.5]$ channels are calculated 
within a $r=1\farcs 5$ aperture.

   All other objects in our samples are either blended with (denoted by ``B''
in Table 1) or severely contaminated (``C'') by foreground neighbors in IRAC
and thus no useful information can be obtained. 

\begin{figure}[tbp]
\centering
\includegraphics[width=\linewidth]{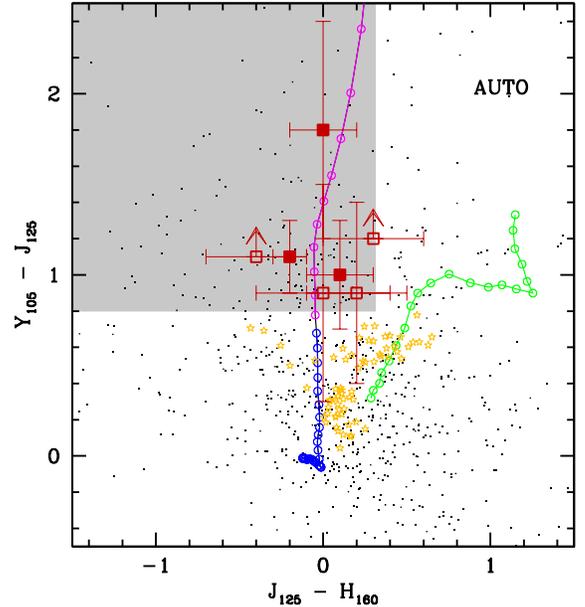} % for two-column
\caption{Selection of $Y_{105}$-dropouts in the $Y_{105}J_{125}H_{160}$
color-color diagram. 
The small dots are field objects that have S/N$<2$ in
ACS. The grey area is our selection region, where the brighter candidates 
($J_{125}\leq 26.2$~mag) are plotted as red filled squares, and the fainter
candidates as red open squares. 
The blue-magenta track shows a young
galaxy (using a BC03 model of age 100~Myr and no dust extinction)
at $z=5$--10, with the blue section for $z<7.7$ and the magenta
section for $z\geq 7.7$. The orange stars show the colors of Galactic brown
dwarfs (Leggett et al. 2002). The green symbols show the colors of a typical
red galaxy at $z\approx 1$--3 simulated using a BC03 model 
($\tau=50$~Myr and age of 2.0~Gyr). 
}
\end{figure}

\begin{figure}[tbp]
\centering
\includegraphics[width=\linewidth]{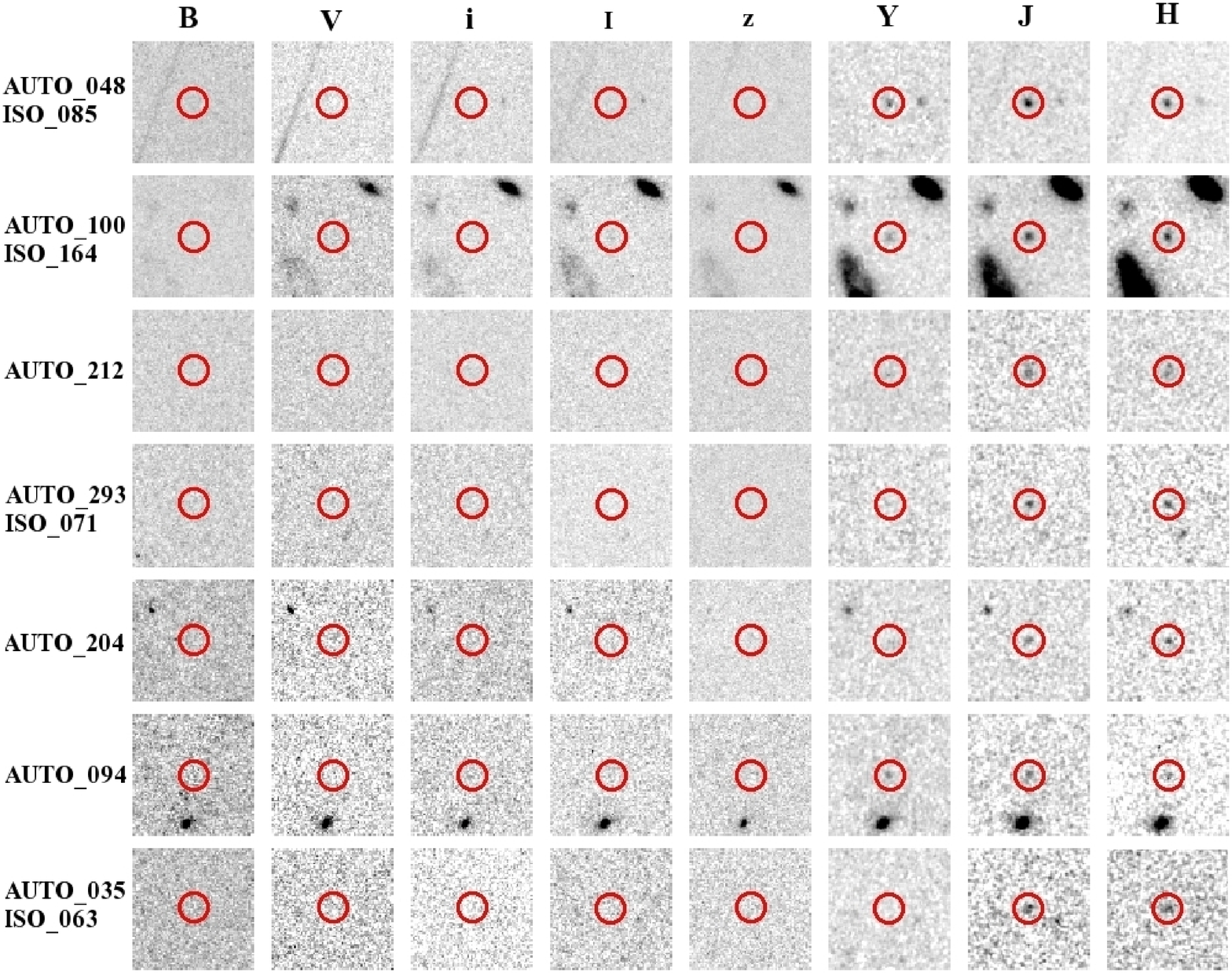} % for two-column
\caption{Images of the $Y_{105}$-dropouts in the AUTO sample, each
$3\farcs 1$ on a side. The locations of the candidates are at the center
marked by red circles, which are $0\farcs 5$ in radius. The common objects
in the comparison ISO sample (see appendix) are also labeled.
}
\end{figure}

\begin{figure}[tbp]
\centering
\includegraphics[width=\linewidth]{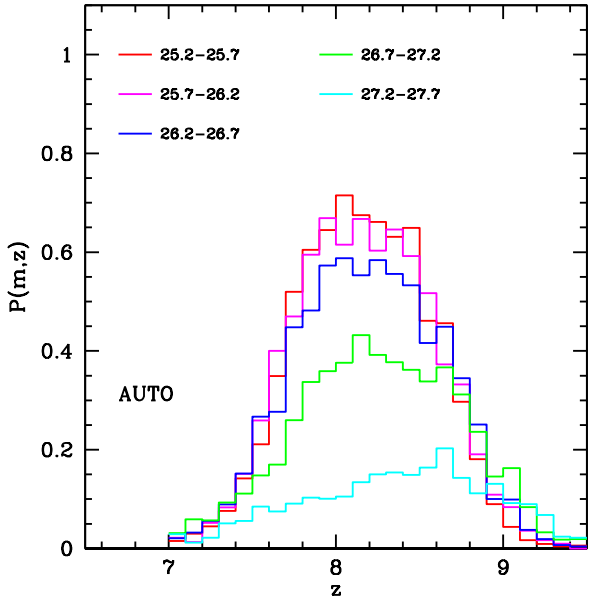} % for two-column
\caption{Redshift selection functions derived from simulations of our color
selection in different magnitude ranges. The labeled magnitude ranges are
based on {\tt MAG\_AUTO} in $J_{125}$.
}
\end{figure}

\subsection{Summary of Sample Limitations}

   Our sample suffer from the same kinds of incompleteness
and contamination that are intrinsic to any dropout samples because of their
selection technique. While a number of these issues have been mentioned in
the above sections, we review them here before 
proceeding to discuss the statistical implications of our samples to
astrophysical questions.

   There are two main factors that contribute to the incompleteness, which
cause significant differences among samples as demonstrated in \S 3.3.1 and
3.3.2. First of all, an object whose intrinsic brightness is above our
sensitivity limit could be undetected because its observed brightness could
fall below the detection threshold due to noise. 
%This is essentially the
%well-known Malmquist bias, with the only difference that the subjects discussed
%here are those sources that are not detected. 
Secondly, a genuine $z\approx 8$
galaxy that is detected in source extraction could still escape our color
selection because it could be ``scattered'' out of the color selection region
(see Figure 1) due
to photometric error, or it could be incorrectly ``vetoed'' because of some
unrelated events within its vicinity (for example, noise spikes) that
accidentally increase its measured S/N in the veto images above the adopted
quantitative veto threshold.  Both effects have been taken into account by 
our simulations, and therefore can be corrected statistically.

    The major factor that causes the contamination is the inclusion of
low-redshift objects with spectral features that could mimic the Lyman-break at
$z\approx 8$. Such contaminators could be selected because their 
$Y_{105}-J_{125}$ colors meet our color threshold, either intrinsically, or
after being boosted by photometric error (i.e., being scattered into the 
selection region). Using the $H_{160}$ data, our secondary $J_{125}-H_{160}$
criterion reduces the number of contaminators by rejecting those that have
red SEDs. For other contaminators, the best way to reject them is to check
their visibility in the veto images. As discussed earlier, the quantitative
criterion of S/N~$<2$ gets rid of most of such objects, and the final step
is the visual inspection. The caveat here is that we implicitly
assume that the veto images are deep enough in detecting the contaminators in
the blue bands, but this is not guaranteed. 
Currently we do not have a reliable way to assess this effect.

   We point out that the rejection of contaminators based on their
visibility in the veto images also introduces incompleteness, and that this
happens both in the quantitative pre-screening (requiring S/N~$<2$) step and in
the last visual inspection step. While the former can be statistically 
corrected by the simulations, the latter cannot.

%   Bearing all the caveats in mind, we stress that our sample is as robust as
%the available data allow. It is not surprising that it could be different
%from others that are constructed using different photometry and procedures, and
%when they are different one sample is not 
%necessarily superior to the other because such differences are caused by the
%similar limitations that they all suffer. 

\section{Discussion}

   Our samples are unique in two aspects. (1) We have two
very bright candidates at $J_{125}\leq 26.0$~mag, which are at least 
$\sim 1$~mag brighter than the brightest $Y_{105}$-dropouts previously found.
They are comparable in brightness to the brightest $Y_{098}$-dropouts reported
by Yan et al. (2011) and Trenti et al. (2011a,b) in their WFC3 parallel
surveys, but could be at higher redshifts because the redshift window of this
current $Y_{105}$-dropout selection is higher than that of the existing 
$Y_{098}$-dropouts. (2) Two other $Y_{105}$-dropouts, which are not among the
brightest in the WFC3 bands,  are securely detected in IRAC, which is in sharp
contrast to previous results where only non-detection in IRAC have been
reported for $Y_{105}$-dropouts.  Here we discuss the implications of both.

\subsection{Stellar Population and Stellar Mass}

   We investigate the stellar populations of the IRAC-detected 
$Y_{105}$-dropouts, {\tt AUTO\_035} and {\tt AUTO\_293}, by
analyzing their spectral energy distributions (SED) through template fitting.
The approach here is similar to that of Finkelstein et al. (2011). We first use
the EAZY code (Brammer et al. 2008) and the photometry of the GOODS ACS \& 
CANDELS WFC3 data to estimate photometric redshifts ($z_{phot}$). The full SED
(including the IRAC photometry) is then fitted to a suite of templates based on
the updated models of BC03 (the so-called ``CB07'' models). A Salpeter initial
mass function (Salpeter 1955) and
the metallicities of 0.02--1$Z_\odot$ are adopted. The templates have a range
of exponentially decreasing and increasing SFHs
(see also Papovich et al. 2011), and can include nebular lines based on the
number of ionizing photons and metallicity of a given model (B. Salmon et al. in
preparation). We assume the extinction law of Calzetti (2001) with
$E(B-V)$$=$0--0.5~mag and the H I absorption as formulated in Madau (1995). The
results are summarized in Figure 4. We note that fitting redshift and other
properties simultaneously does not signficantly change these results,
and the differences are captured in the errors that we quote here.

    The best-fit photometric redshifts are $8.5^{+0.2}_{-4.5}$ and 
$8.9^{+0.2}_{-0.5}$ for {\tt AUTO\_035} and {\tt 293}, respectively, which are
consistent with the redshift window of our color selection. Adding the formal
$I_{814}$ limit to the fitting process does not change the results.  We fit the
SEDs with and without the contribution from nebular lines, and both results
suggest high stellar mass ($\mathcal M_*$) in the range of $10^{9}$ to 
$10^{10} M_\odot$ for both objects. Including nebular emission lines, we obtain
$2.5^{+4.7}_{-1.4}\times 10^{9}$ and 
$0.9^{+3.0}_{-0.1}\times 10^{9}$~$M_\odot$ for {\tt AUTO\_035} and {\tt 293}
respectively, while without nebular emission lines these are
$10.5^{+1.8}_{-8.1}\times 10^{9}$ and 
$2.6^{+1.8}_{-1.4}\times 10^{9}$~$M_\odot$, respectively.
The error bars reflect the 68\% confidence level of the fit.
Labb\'e et al. (2010a) derived the stellar masses of galaxies at lower
redshifts of $z\approx 7$ using models without nebular emission, and the 
average value is
$\sim 10^9$~$M_\odot$. The two objects in our sample have comparable or
even larger values using similar models, which are surprisingly high at 
such an early epoch. 

    Using the $\mathcal M_*$ estimates, we obtain the first measurement of the
mass function (MF) of galaxies at $z\approx 8$ at the high-mass end. As we only
have two objects whose $\mathcal M_*$ values are reasonably similar, we opt to
count them within one mass bin of a $\pm 0.4$~dex bin size. The two different
sets of models result in significantly different mass estimates, and therefore
the bin center is different for each case. Assuming a top-hat selection
function of our survey within $7.7\leq z\leq 8.7$, we get
$\phi(log\frac{M_*}{M_\odot})|_{9.2\pm 0.4} = (1.9^{+2.4}_{-0.6})\times 10^{-5}$~Mpc$^{-3}$~dex$^{-1}$ if we adopt the values derived using the models with
nebular emission. 
If we apply the effective volume ($V_{eff}$) correction of the survey
instead (see \S 3.4), we obtain 
$(3.1^{+4.0}_{-1.0})\times 10^{-5}$~Mpc$^{-3}$~dex$^{-1}$.
Here we use
$V_{eff}=\int V_{eff}(m_J)dm_J=\int\int dm_JdzP(m_J,z)dV/dz$, where $dzdV/dz$ is the unit co-moving volume
at redshift $z$, and $P(m_J,z)$ is the redshift selection function 
at different magnitudes $m_J$ as derived through
simulations (see Figure 3). The error bars here reflect the 68\% interval of the
uncertainties caused by the Poisson noise in the sample.
For the case of using the $\mathcal M_*$ estimates based on the models without
nebular lines, all the above values are applicable at 
$\phi(log\frac{M_*}{M_\odot})|_{9.7\pm 0.4}$.

   The contribution of these objects to the global stellar mass density, 
$\rho_{*}$, at $z\approx 8$ within our survey volume is 
$(3.3^{+7.4}_{-1.4}) \times 10^{4}$ and 
$(12.6^{+3.4}_{-9.0}) \times 10^{4}$~$M_\odot$~Mpc$^{-3}$ based on the models
with and without nebular emission lines, respectively, if assuming a top-hat 
selection function at $7.7\leq z \leq 8.7$. 
The errors reflect the 68\% confidence level of the fitting results.
If we apply the same correction for $V_{eff}$ as above, we obtain
$\rho_{*}=(5.5^{+12.6}_{-2.4}) \times 10^{4}$ and
$(21.2^{+5.8}_{-15.3}) \times 10^{4}$~$M_\odot$~Mpc$^{-3}$.

   We derive the mass-to-light ratio of these two objects using their 
4.5~$\mu$m flux as a proxy to $L_V$. Using the fit results with contribution
from nebular lines, we obtain
$\mathcal M_*/L_V=0.07^{+0.14}_{-0.04}$ and $0.04^{+0.13}_{-0.00}$ for
{\tt AUTO\_035} and {\tt 293}, respectively.
Without the contribution from nebular lines, we obtain
$\mathcal M_*/L_V=0.31^{+0.06}_{-0.24}$ and $0.12^{+0.08}_{-0.06}$.
The latter values are in general agreement with those obtained by Labb\'e
et al. (2010a,b) through stacking analysis of the IRAC data of the 
$Y_{098}$-dropouts and the $Y_{105}$-dropouts in the WFC3 ERS field (see also
Gonz\'alez et al. 2011) and the HUDF using similar models without nebular lines.
On the other hand, Labb\'e et al. (2010b) estimated that nebular lines have a
small impact on $\mathcal M_*/L_V$, reducing it by $\sim 0.2$~dex. Our results
suggest a stronger effect, reducing $\mathcal M_*/L_V$ by $\sim 0.5$--0.6 dex.

\begin{figure}[tbp]
\centering
\includegraphics[width=\linewidth]{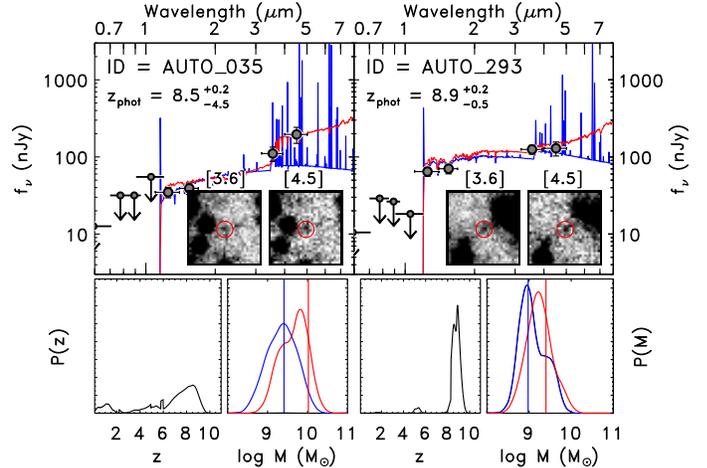}
\caption{Summary of SED fitting results for {\tt AUTO\_035} (left) and
{\tt 293} (right). 
The results obtained from the fit with and without the
contributions from nebular lines are coded in blue and red, respectively. The
top panels show the observed SEDs (grey circles) and the best-fit templates,
and the insets display their 3.6 and 4.5~$\mu$m image ($18\farcs 6
\times 18\farcs 6$). The bottom panels show the likelihood functions of
$z_{phot}$ and $M_*$, and the best-fit values are indicated by the vertical
lines. 
}
\end{figure}

\subsection{Constraint on the LF}

   Here we use our sample to constrain the LF at $z\approx 8$. We adopt the
$J_{125}$ {\tt MAG\_AUTO} values of the candidates as their total magnitudes. 

   After applying the correction for $V_{eff}$, we obtain the stepwise LF in
the five 0.5-mag bins (25.45, 25.95, 26.45, 26.95, 27.45)~mag as 
$\phi = (2.5^{+5.8}_{-0.6}, 2.5^{+5.8}_{-0.6}, 2.8^{+6.3}_{-0.6}, 3.6^{+8.3}_{-0.8}, 24.2^{+22.6}_{-5.6})\times 10^{-5}$~Mpc$^{-3}$~mag$^{-1}$.
These number densities are shown in Figure 5,
and compared to the predictions from a number of
Schechter LF estimates at $z\approx 8$ (M10; Y10; B11; L11).
The black squares are our observed densities, while the red squares are the
densities after correcting for $V_{eff}$. The error bars represent a Bayesian
68\% credible interval, indicating the central 68 percentile range for the
posterior distribution of the true number density assuming Poisson statistics.
These uncertainties account for the incompleteness, but do not account for
any systematic uncertainties from the contamination due to low-z interlopers.
For ease of direct comparison to observations, 
surface density and apparent magnitude scales are also provided on the same
figures to present these results in terms of differential number densities
versus apparent $J_{125}$ magnitudes.
For comparision, we also plot the {\it observed} densities (i.e., before 
corrections for their corresponding $V_{eff}$ values)
extracted from the sample of Y10 in the
HUDF09 proper and those of B11 in the HUDF09, HUDF09P1 and 
HUDF09P2 {\it after} applying the additional criteria of 
$Y_{105}-J_{125}\geq 0.8$~mag and $J_{125}-H_{160}\leq 0.3$~mag.

\begin{figure}[tbp]
\centering
\includegraphics[width=\linewidth]{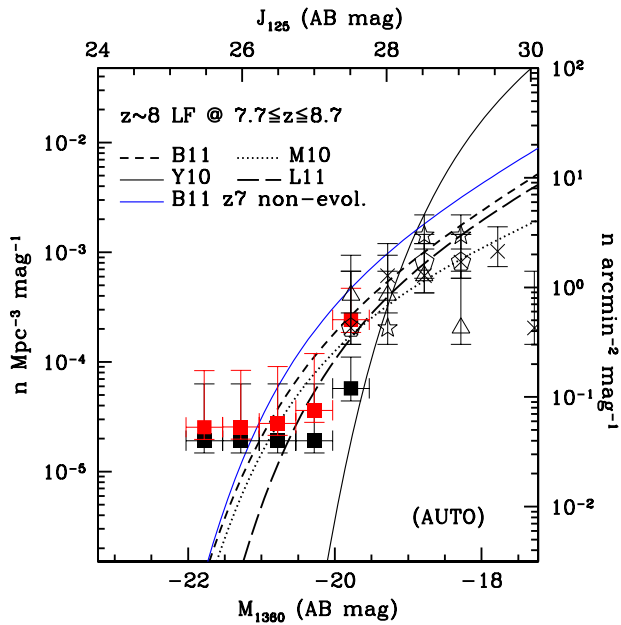}
\caption{Constraints on the very bright-end of the LF at $z\approx 8$ based on
the sample of $Y_{105}$-dropouts in CANDELS/Deep from this current work.
The black squares are the differential surface densities inferred from the
number counts in our sample, while the red squares are those after correction
for $V_{eff}$ with respect to the volume within $7.7\leq z\leq 8.7$. The count
predictions over $7.7\leq z\leq 8.7$ are from the Schechter LF estimates at 
$z\approx 8$ from B11, M10, Y10, and L11 are shown as various
black curves, together with a non-evolution one at $z\approx 7$ from B11
as the blue curve. For comparison, different symbols at the
faint-end show the raw (observed) surface densities of $Y_{105}$-dropouts based
on the following samples: Y10 (HUDF -- stars), B11 (HUDF -- crosses; HUDFP1 --
pentagons; HUDFP2 -- triangles).
} 
\end{figure}

   An intriguing feature that our samples reveal is that there could be an
excess of bright objects at $J_{125}\lesssim 26.2$~mag with respect to any of
the existing Schechter LF estimates at $z\approx 8$ in the literature would
predict. The inferred number density at this bright end is even higher than the
prediction from a non-evolution $z\approx 7$ LF (e.g., the one from B11). This
excess is still evident with the observed counts {\it before} applying the
incompleteness correction. The observed counts at $J_{125}\lesssim 26.2$~mag in
Figure 5 comes from objects {\tt AUTO\_048} and {\tt AUTO\_100}.
Based on their photometry in $J_{125}$ and assuming $z=8$, they
have $M_{UV}$ of $-21.54$ and $-21.21$, respectively.
While such high luminosities have been observed in the spectroscopically
confirmed $z\approx 7$ galaxies of Ono et al. (2012), they are 5--9$\times$
more luminous than any previously known $Y_{105}$-dropouts. 
Such an excess
is unlikely due to cosmic variance, whose effect is expected to be small
compared to the Poisson errors for our observed number density if we use the
formalism of Trenti \& Stiavelli (2008).

   We note that, however, that O12 do not see such an excess, even though their
sample also includes our brightest candidate 
{\tt AUTO\_048}. The main reason for this discrepancy could be that
O12 use a larger volume in the calculation. They have included part of the
CANDELS Wide field in the GOODS-S region, which is a major contributor to their
larger volume. However, this area does not have sufficiently deep $Y_{105}$
data to apply our selection criterion of $Y_{105}-J_{125}\geq 0.80$~mag down to
$J_{125}\approx 26.0$~mag. O12 have adopted a less stringent criterion of
$Y_{105}-J_{125}\geq 0.45$~mag, which allows them to construct a 
$Y_{105}$-dropout sample in this area at the cost of including candidates at
lower redshifts ($z\approx 7.2$). The advantage of their approach is that they
can incorporate the WFC3 ERS field in the GOODS-S region as well, which 
further increases their survey volume. While the WFC3 ERS field has $Y_{098}$
but not $Y_{105}$ data, the redshift range probed by their $Y_{098}$-dropouts
in this area overlaps that of their $Y_{105}$-dropouts elsewhere, and therefore
O12 has combined the $Y_{098}$- and the $Y_{105}$-dropouts to address the LF
averaged over a wide redshift range ($z\approx 7$--8.7).

   In contrast, our color criteria select galaxies in a higher range of
redshifts, $7.7\lesssim z\lesssim 8.7$, and the bright-end excess (with respect
to a smooth Schechter function) that we find suggests that our understanding
of the LF at $z\approx 8$ is still far from being complete. The significance of
this excess is still subject to small number statistics, and we cannot rule out
the possibility that one or both of our brightest objects could be interlopers
at low redshifts. However, we should
point out that similar bright-end excess at $z\sim 8$ has also been 
suggested in other studies, for example in the bright $Y_{098}$-dropout
study in HIPPIES (Yan et al. 2011) at similar depths, in the ground-based
survey of Laporte et al. (2011) at a brighter level, and in the much wider
survey in Capak et al. (2011) at an even brighter limit
\footnote{While Bowler et al. (2012) derived different $z_{ph}$ ($\sim 2$) for
the candidates in Capak et al. (2011), the true nature of these objects still
remains uncertain.}. 

\subsection{Implications for Star Formation in the Early Universe }

   Our $z\approx 8$ candidate galaxy samples have two very luminous objects and
two high stellar mass objects. We have presented tentative evidence that the
bright end of the galaxy LF at $z\approx 8$ might not follow the exponential
cut-off of the Schechter function. While the stellar mass 
function at this redshift has not yet been determined, we have provided the
first, albeit still tentative measurement at the high-mass end and shown that
some galaxies with stellar masses as high as a few $\times 10^9$~$M_\odot$ might
already in place at $z\approx 8$.

    The physical origin of the exponential cut-off at the bright end of the 
LF or the high mass end of the MF seen at lower redshifts remains a matter of
debate. At $z\sim 0$, star formation is ``quenched'' in halos with masses
greater than $\sim 10^{12}$~$M_\odot$ (``quenching mass''), perhaps by AGN 
feedback (e.g., Croton et al. 2006; Somerville et al. 2008; Gabor et al. 2011).
However, at higher redshift ($z\lesssim 2$), the quenching mass may be higher
(Dekel et al. 2009; Behroozi et al. 2010), and the existence of massive, 
rapidly star forming galaxies at $z>2$ is well established. At $z\sim 2$, the
observed exponential cut-off in the rest-frame UV LF appears to be due to dust 
(Reddy et al. 2010) rather than quenching. At very high redshifts, it has been
noted before that the very blue rest-frame UV colors of $z\approx 7$ candidate 
galaxies (some of them confirmed) suggest that they may contain little
dust (e.g., Oesch et al. 2010; M10; Y10; Bunker et al. 2010). Thus, at
$z\approx 8$, when (1) the quenching mass is much higher than the typical halo
mass, and (2) dust has little effect, perhaps we should not expect to see an
exponential cutoff in the luminosity or mass functions of UV-bright galaxies.

   It is interesting that the estimated stellar masses and number densities
could imply a rather high efficiency of conversion of baryons into stars. For
the lower stellar mass estimates, which are obtained using the models with
nebular emission (the average is $\sim 2\times 10^9$~$M_\odot$), and assuming
that $\sim 20$\% of the available baryons have been converted to stars, the
implied halo masses are on the order of $\sim 10^{11}$~$M_\odot$, for which the
expected number density of dark matter halos 
in the currently favored LCDM cosmology is comparable to the observed number 
density of objects. This is consistent with the predicted star formation
efficiencies and host halo masses at $z\approx 8$ from cosmological 
hydrodynamic simulations. 
However, if the true stellar masses are higher, or the star formation 
efficiency is lower, a rapidly growing tension arises between the number 
density of dark matter halos and the observed number density of galaxies 
(above halo mass $\sim 10^{11}$~$M_\odot$, the halo number density declines by
about two orders of magnitude for a factor of $\sim 3$ increase in mass). 
While converting 20\% of the available baryons into stars may not sound 
excessive, this is in fact the \emph{maximum} value that has been inferred at
any epoch. Due to the presumably very low metallicity of the gas in these early
objects, we might have expected much lower star formation efficiencies than 
are seen locally (e.g. Krumholz \& Dekel 2010). 

\section{Summary}

   In this work, we search for candidate galaxies at $z\approx 8$ in the
CANDELS Deep GOODS-S field and study their properties. Our sample of 
$Y_{105}$-dropouts hints that the number density of $z\approx 8$ galaxies
at the bright-end might be higher than expected from the previous Schechter LF 
estimates, which lends support to the suggestion made by a number of earlier
studies that there could be a bright-end excess in the galaxy number density at
very high redshifts. Furthermore, two of our candidates are securely
detected at 3.6 and 4.5~$\mu$m in {\it Spitzer} data. These are the first 
$Y_{105}$-dropouts individually detected at these wavelengths. Their derived
stellar masses are on the order of $\sim 10^9$~$M_\odot$, from which we obtain
the first measurement of the high-mass end of the galaxy stellar mass function
at $z\approx 8$. If the high number densities of very luminous and very massive
galaxies at $z\approx 8$ are real, they could imply a large stellar-to-halo mass
ratio and an efficient conversion of baryons to stars at very early time in 
the cosmic stellar mass assembly history.

\acknowledgements

We thank the referee for the useful comments, which improved the quality of
this work. We also thank B. Mobasher and J. Dunlop for their comments
on an earlier version of this paper.
H.Y. acknowledges the support of NASA grant HST-GO-11192.1.
Support for Program number HST-GO-12060 was provided by NASA through a grant
from the Space Telescope Science Institute, which is operated by the
Association of Universities for Research in Astronomy, Incorporated, under
NASA contract NAS5-26555.

\begin{deluxetable}{cccccccl}
%\rotate
\tablewidth{0pt}
\tablecolumns{8}
\tabletypesize{\scriptsize}
\tablecaption{{\tt MAG\_AUTO}-selected $Y_{105}$-dropouts in CANDELS GOODS-S Deep Region}
\tablehead{
\colhead{ID} &
\colhead{RA \& DEC (J2000)} &
\colhead{$J_{125}$\tablenotemark} &
\colhead{$Y_{105}-J_{125}$\tablenotemark{a}} &
\colhead{$J_{125}-H_{160}$\tablenotemark} &
\colhead{[3.6]\tablenotemark{b}} &
\colhead{[4.5]\tablenotemark{b}} &
\colhead{Other ID\tablenotemark{c}} 
}
\startdata
   AUTO\_048 & 3:32:49.936 $-$27:48:18.101 & 25.65$\pm$0.07 & 1.1$\pm$0.2 & $-$0.2$\pm$0.1 &       C        &       C        & ISO\_085; O12-2499448181   \\
   AUTO\_100 & 3:32:41.417 $-$27:44:37.831 & 25.98$\pm$0.14 & 1.8$\pm$0.6 &    0.0$\pm$0.2 &       B        &       B        & ISO\_164                   \\
   AUTO\_212 & 3:32:20.965 $-$27:51:37.073 & 26.39$\pm$0.15 & 1.0$\pm$0.3 &    0.1$\pm$0.2 &    $>26.8$     &   $>26.2$      & O12-2209751370             \\
   AUTO\_293 & 3:32:20.981 $-$27:48:53.467 & 26.88$\pm$0.17 &   $>1.1$    & $-$0.4$\pm$0.3 & 26.16$\pm$0.13 & 26.12$\pm$0.22 & ISO\_071; O12-2209848535   \\
   AUTO\_204 & 3:32:18.185 $-$27:52:45.566 & 27.38$\pm$0.19 & 0.9$\pm$0.5 &    0.2$\pm$0.3 &    $>26.7$     &   $>26.2$      & O12-2181852456             \\
   AUTO\_094 & 3:32:40.675 $-$27:45:11.624 & 27.39$\pm$0.27 & 0.9$\pm$0.6 &    0.0$\pm$0.4 &       C        &       C                                     \\
   AUTO\_035 & 3:32:34.998 $-$27:49:21.623 & 27.55$\pm$0.21 &   $>1.2$    &    0.3$\pm$0.3 & 26.29$\pm$0.22 & 25.67$\pm$0.25 & ISO\_063; O12-2350049216   \\
\enddata

\tablenotetext{a.}{For the objects that are not detected in $Y_{105}$,
  their $Y_{105}-J_{125}$ limits are caculated
  using the 2-$\sigma$ flux upper limits in $Y_{105}$ as measured within the
  {\tt MAG\_AUTO} apertures.}
\tablenotetext{b.}{The IRAC magnitudes are {\tt MAG\_APER} within $r=1\farcs 5$
  aperture, and limits are based on 2-$\sigma$ flux upper limits within the
  same aperture, both of which are with an aperture correction to a total flux.
  ``B'' or ``C'' means that the object is blended with or contaminated by
  foreground neighbor(s) in IRAC.}
\tablenotetext{c.}{For the common candidates in the AUTO and the ISO
  samples (see appendix), their ID's in the other set are given. In addition,
  if an object is also in the O12 sample, its ID in O12 is given as well.}

\end{deluxetable}

\newpage
\centerline{APPENDIX}

   Due to the nature of the dropout selection technique, it is not unusual that
samples constructed by independent groups can differ significantly
on an object-by-object basis (see \S 3). To further demonstrate this point, we
constructed a separate $Y_{105}$-dropout sample to mimic an independent study.

   We used exactly the same data as in \S 2, and therefore any difference in
this sample should be attributed to the selection technique. We followed the
same procedures as in \S 3, but used {\tt MAG\_ISO} instead when calculating
colors. The motivation of adopting {\tt MAG\_ISO} in this exercise is that 
isophotal apertures are commonly used in measuring Lyman-break galaxy colors
(e.g., Steidel et al. 2003) because they often lead to higher $S/N$, even if
they do not necessarily include all of the light from each galaxy.
Hereafter we refer to this sample as the ``ISO sample'' to separate from
the one described in the main text, which we refer to as the ``AUTO sample''.
The initial candidates in the ISO sample were also visually examined by the
same group of six inspectors. The visual inspection of this sample was done
about one month later than that for the AUTO sample, and therefore even when
we were examining the same object that is also in the AUTO sample we did not
keep the memory of the result from the last time, and hence we mimicked an 
``independent'' study as much as we could.

   The final ISO sample consists of eleven candidates. Fig. 6 shows their
selection in the $Y_{105}J_{125}H_{160}$ color space, while Fig. 7 shows the
redshift selection function. Images of these candidates are displayed in Fig. 8.
Table 2 lists their photometric information.

\begin{figure}[tbp]
\centering
\includegraphics[width=6.0cm]{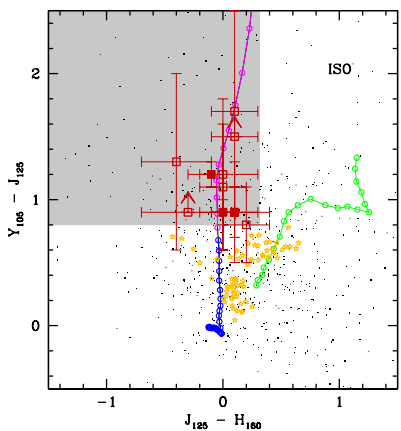} % for two-column
\caption{Similar to Figure 1, but for selection of $Y_{105}$-dropouts using
{\tt MAG\_ISO} magnitudes.
}
\end{figure}

\begin{figure}[tbp]
\centering
\includegraphics[width=6.0cm]{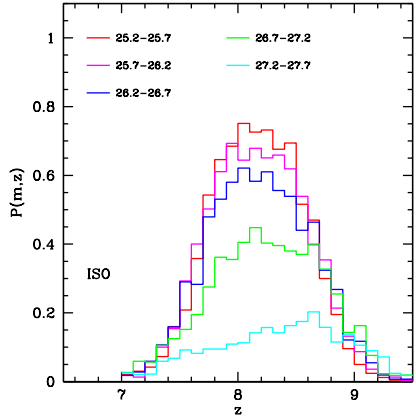} % for two-column
\caption{Similar to Figure 2, but for the ISO $Y_{105}$-dropout sample.
The labeled magnitude ranges are still based on {\tt MAG\_AUTO} in $J_{125}$,
as the simulation only has control over the input total magnitudes, which are
best represented by {\tt MAG\_AUTO}. To be consistent, the correction of
$V_{eff}(m)$ for the ISO sample is done at the corresponding {\tt MAG\_AUTO}
ranges.
}
\end{figure}

\begin{figure}[tbp]
\centering
\includegraphics[width=8.0cm]{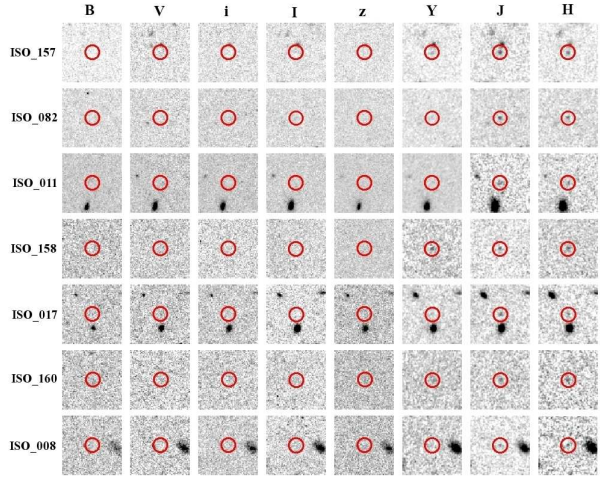} % for two-column
\caption{Similar to Fig. 2, but for the ISO sample. The ones that are in
common with the AUTO sample are not displayed here.
}
\end{figure}

   For the sake of completeness, we also compared the ISO sample to the O12's
sample. Based on the photometry reported in O12, nine of their candidates are
expected to meet our color criteria (see also \S 3.3.3). However, only three of
them are included in our ISO sample (see Table 2). Among the five that
are not in our sample, O12's CANDY-2432246169 has $S/N<5$ in $J_{125}$ and
$H_{160}$ and is not included in our {\tt MAG\_ISO} catalog. CANDY-2272447364
has $S/N>2$ in $V_{606}$ and does not satisfy our optical non-detection
criterion. CANDY-2277945141 and 2320345371 have $Y_{105}-J_{125}=0.74$ and
0.57~mag based on our photometry and hence do not meet out color criteria. 
The other two, O12's CANDY-2209751370 and 2181852456, meet all our color
criteria, but were rejected during the visual inspection when constructing our
ISO sample. Note that these latter two objects actually are in our AUTO
sample ({\tt AUTO\_212} and {\tt 204}, respectively) and
survived the visual inspection for our AUTO sample (see below).

   A more constructive comparision is that between this ISO sample and the 
AUTO sample, which shows that they have four objects in common (see Table 2).
There are seven ISO candidates not in the AUTO sample. Five of them were
rejected in the AUTO selection because they have S/N~$>2$ within their 
{\tt MAG\_AUTO} apertures in at least one veto image ({\tt ISO\_011, 017, 157,
158, 160}). One was rejected because its {\tt MAG\_AUTO} $Y-J$ color limit does
not satisfy $Y-J>0.8$~mag ({\tt ISO\_008}). The other one ({\tt ISO\_082})
satisfies all the quantitative criteria in the AUTO selection, however it was
rejected in the visual inspection step during the AUTO selection run.

   On the other hand, four AUTO candidates are not in the ISO sample. Two of
them were rejected because they have {\tt MAG\_AUTO} $Y-J$ color or limit below
the $Y-J>0.8$~mag threshold ({\tt AUTO\_094 and 368}). The other two satisfy 
all the quantitative criteria in the ISO selection, however they were rejected
in the visual inspection step during the ISO selection run
({\tt AUTO\_204 and 212}).

   This internal comparison of our two samples thus further demonstrates the
points addressed earlier: (1) adoption of different photometry result in
samples that can be significantly different on an object-by-object basis, and
(2) visual inspections at different time, especially the inspections of the
veto images, can also result in differences in this sense because such 
inspections are not guaranteed to be fully repeatable for a large sample at the
S/N$<2$ level.

   Nevertheless, the main statistical trends revealed by the ISO sample are very
similar to those inferred from the AUTO sample. Both samples include the
same two bright $Y_{105}$-dropouts ({\tt AUTO\_048/ISO\_085} and
{\tt AUTO\_100/ISO\_184}), and the same two IRAC-detected ones
({\tt AUTO\_293/ISO\_071} and {\tt AUTO\_035/ISO\_063}). For completeness,
Fig. 9 shows the constraint from the ISO sample on the LF, where one can
see that the bright-end excess is still present. Similarly, the constraints on
the MF and the stellar mass density derived from the ISO sample agree with
those based the AUTO sample to within $\lesssim 10$\%, and the small 
differences are mainly caused by the slightly different $V_{eff}$ corrections
over the range of interest. From this exercise, we believe that the main
results presents in this work are robust.

\begin{figure}[tbp]
\centering
\includegraphics[width=7.0cm]{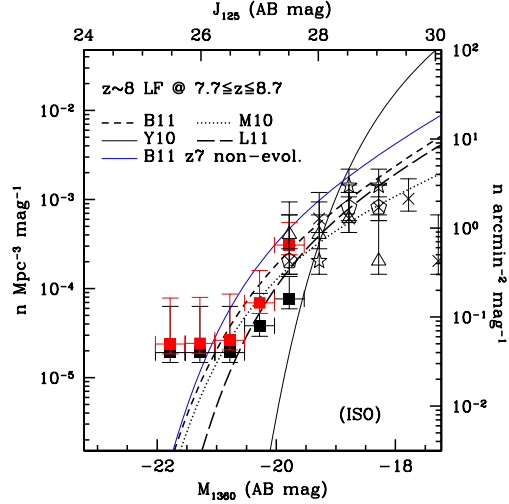}
\caption{Same as Figure 5, but for the ISO sample. Note that the $J_{125}$
magnitudes shown here are the {\tt MAG\_AUTO} magnitudes, because these are
taken as the total magnitudes. 
} 
\end{figure}

\begin{deluxetable}{cccccccl}
%\rotate
\tablewidth{0pt}
\tablecolumns{8}
\tabletypesize{\scriptsize}
\tablecaption{{\tt MAG\_ISO}-selected $Y_{105}$-dropouts in CANDELS GOODS-S Deep Region}
\tablehead{
\colhead{ID} &
\colhead{RA \& DEC (J2000)} &
\colhead{$J_{125}$\tablenotemark{a}} &
\colhead{$Y_{105}-J_{125}$\tablenotemark{b}} &
\colhead{$J_{125}-H_{160}$\tablenotemark{c}} &
\colhead{[3.6]\tablenotemark{d}} &
\colhead{[4.5]\tablenotemark{d}} &
\colhead{Other ID\tablenotemark{e}} 
}
\startdata
   ISO\_085  & 3:32:49.936 $-$27:48:18.100 & 25.65$\pm$0.07 & 1.2$\pm$0.1 & $-$0.1$\pm$0.1 &       C        &       C        &  AUTO\_048; O12-2499448181 \\
   ISO\_164  & 3:32:41.417 $-$27:44:37.831 & 25.98$\pm$0.14 & 0.9$\pm$0.1 &    0.0$\pm$0.1 &       C        &       C        &  AUTO\_100                \\
   ISO\_157  & 3:32:42.882 $-$27:45:04.268 & 26.50$\pm$0.13 & 0.9$\pm$0.2 &    0.1$\pm$0.2 &       B        &       B        &                           \\
   ISO\_071  & 3:32:20.981 $-$27:48:53.468 & 26.88$\pm$0.17 & 1.7$\pm$0.8 &    0.1$\pm$0.2 & 26.16$\pm$0.13 & 26.12$\pm$0.22 &  AUTO\_293; O12-2209848535 \\
   ISO\_082  & 3:32:14.133 $-$27:48:28.911 & 27.16$\pm$0.19 & 1.1$\pm$0.5 &    0.0$\pm$0.2 &    $>27.1$     &    $>26.5$     &                           \\
   ISO\_011  & 3:32:14.469 $-$27:51:48.542 & 27.24$\pm$0.30 & 1.2$\pm$0.6 &    0.0$\pm$0.3 &       B        &       B        &                           \\
   ISO\_158  & 3:32:47.953 $-$27:44:50.436 & 27.28$\pm$0.23 & 0.8$\pm$0.3 &    0.2$\pm$0.2 &    $>26.7$     &    $>25.9$     &                           \\
   ISO\_063  & 3:32:34.999 $-$27:49:21.622 & 27.55$\pm$0.21 &    $>1.5$   &    0.1$\pm$0.2 & 26.29$\pm$0.22 & 25.67$\pm$0.25 &  AUTO\_035; O12-2350049216 \\
   ISO\_017  & 3:32:18.091 $-$27:51:18.492 & 27.57$\pm$0.18 & 1.3$\pm$0.7 & $-$0.4$\pm$0.3 &       B        &       B        &                           \\
   ISO\_160  & 3:32:46.111 $-$27:44:47.997 & 27.88$\pm$0.50 & 0.9$\pm$0.4 &    0.1$\pm$0.3 &    $>26.7$     &    $>26.1$     &                           \\
   ISO\_008  & 3:32:16.915 $-$27:52:01.878 & 28.00$\pm$0.27 &    $>0.9$   & $-$0.3$\pm$0.4 &    $>26.7$     &    $>26.2$     &                           \\
\enddata

\tablenotetext{a.}{The quoted total magnitudes in $J_{125}$ are their 
  {\tt MAG\_AUTO} values.}
\tablenotetext{b.}{The $Y_{105}-J_{125}$ colors are based on their 
  {\tt MAG\_ISO} magnitudes. For the objects that are not detected in 
  $Y_{105}$, their $Y_{105}-J_{125}$ limits are caculated using the 2-$\sigma$
  flux upper limits in $Y_{105}$ as measured within the  {\tt MAG\_ISO}
  apertures.}
\tablenotetext{c.}{The $J_{125}-H_{160}$ colors are based on their 
  {\tt MAG\_ISO} magnitudes.}
\tablenotetext{d.}{The IRAC magnitudes are {\tt MAG\_APER} within $r=1\farcs 5$
  aperture, and limits are based on 2-$\sigma$ flux upper limits within the
  same aperture, both of which are with an aperture correction to a total flux.
  ``B'' or ``C'' means that the object is blended with or contaminated by
  foreground neighbor(s) in IRAC.}
\tablenotetext{e.}{For the common candidates in the AUTO and the ISO
  samples, their ID's in the other set are given. In addition, if an object
  is also in the O12 sample, its ID in O12 is given as well.}

\end{deluxetable}

\end{document}